\documentclass[preprint,aps,floatfix,nofootinbib]{revtex4}
\usepackage{graphics}
\usepackage{graphicx}
\usepackage{amssymb}		
\usepackage{amsmath}
\usepackage{verbatim}
\usepackage{pstricks}
\usepackage{slashed}

\newcommand{\beq}{\begin{equation}}
\newcommand{\eeq}{\end{equation}}
\newcommand{\bea}{\begin{eqnarray}}
\newcommand{\eea}{\end{eqnarray}}

\newcommand{\Eq}[1]{{\rm Eq.~(\ref{#1})}}

\newcommand{\roughly}[1]%
    {{\mathrel{\raise.3ex\hbox{$#1$\kern-.75em\lower1ex\hbox{$\sim$}}}}}
\newcommand{\lsim}{\mathrel{\roughly<}}

\newcommand{\be}{\ensuremath{\beta}}

\newcommand{\ka}{\ensuremath{\kappa}}

\begin{document}

\title{Higgsstrahlung from R-hadrons}

\author{Markus A. Luty}
\vspace{0.2cm}
\affiliation{Physics Department, University of California at Davis, Davis, CA 95616}

\author{Daniel J. Phalen}
\vspace{0.2cm}
\affiliation{Physics Department, University of California at Davis, Davis, CA 95616}

\date{\today}

\begin{abstract}
If $R$ hadrons are discovered at the LHC, investigation of their properties
will be of paramount importance.
One important question is how much of the $R$-hadron mass is due to electroweak
symmetry breaking, {\it i.e.}\ the coupling of the Higgs to $R$-hadrons.
In this paper we show that in models where the Higgs has a sizable coupling to
$R$-hadrons we can readily observe Higgs production in association with
a pair of $R$-hadrons (``Higgsstrahlung'').
This process can be used to distinguish between different models of $R$-hadrons.
It may be the discovery mode of the Higgs for low mass Higgs bosons, 
and provides a low-background Higgs sample
to study $h \to \bar{b}b$.
\end{abstract}

\maketitle

\setcounter{equation}{0}

\section{Introduction}

If there are heavy colored particles that are stable on collider scales,
they will form heavy ``$R$-hadrons.''
These will often be electrically charged, and can therefore be observed at
colliders as ionizing tracks.
Searches for this signal at the Large Hadron Collider (LHC) have already been carried out by
the CMS collaboration with 3.1 pb$^{-1}$ of data \cite{Khachatryan:2011ts}.
If $R$-hadrons are discovered, measuring their detailed properties will
be of paramount importance.
One key property is how much of the mass of the $R$-hadron arises from
the coupling to the Higgs field.
In this paper we consider associated production of the Higgs with a
pair of $R$-hadrons, or ``Higgsstrahlung'' from $R$-hadrons.
We show that this process with $h \to \bar{b}b$
is readily observable at the LHC if the Higgs coupling to $R$-hadrons
is order 1.
This can be used to distinguish among a number of well-motivated
scenarios that lead to $R$-hadrons.
For example, if the $R$-hadron is made of a stop LSP 
or new heavy colored matter fields
we expect sizable Higgs couplings,
while if it is a gluino \cite{Raby:1997pb,Baer:1998pg,Raby:1997ba,Mafi:1999dg,Mafi:2000kg,ArkaniHamed:2004fb} we do not.

This paper is organized as follows.
In Section 2 we motivate and describe
the model we will use as a benchmark for our study.
In Section 3, we describe the search strategy.
Section 4 gives our conclusions.

\section{Benchmark Model}
We now describe a specific model that motivates the signal we will study.
We assume that the hierarchy problem is solved by supersymmetry (SUSY).
The main problem in SUSY is the Higgs mass is generally
too light without fine tuning.
In the minimal supersymmetric Standard Model (MSSM) the Higgs mass can be raised above the tree-level bound of
$m_Z \cos(2\beta)$ by radiative corrections from the stop sector,
but this is generally fine-tuned.
One simple way to address this is to assume the existence of additional
matter fields with large Higgs couplings.
These can raise the Higgs mass with reduced fine-tuning
\cite{Moroi:1991mg,Moroi:1992zk,Babu:2008ge,Babu:2004xg,Martin:2009bg}.
Gauge coupling unification suggests that these extra particles
come in complete $SU(5)$ representations, so some of them will be colored.
These new colored particles must have highly suppressed mixing
to ordinary quarks with the same gauge quantum numbers,
which can be made natural with an approximate $\mathbb{Z}_2$ symmetry 
under which the new particles are odd.
If the new colored particles are stable on cosmological time scales,
they are ruled out by direct searches for heavy stable particles
in sea water \cite{Verkerk:1991jf,Hemmick:1989ns}.%
\footnote{These bounds do not apply if the universe has a low reheat
temperature, so that the new particles are not thermally produced.
Such a scenario requires low-temperature mechanisms to explain the
baryon asymmetry and origin of dark matter.}
However, the $\mathbb{Z}_2$ symmetry may be broken by effects suppressed
by high scales such as the GUT scale, in which case the new colored
particles will be stable on collider scales without cosmological
problems.

As a specific example, we use the QUE model described in
Ref.~\cite{Martin:2009bg}.
The new fields come in a $\bf{10} \oplus \overline{\bf{10}}$ of $SU(5)$
and are denoted $Q, U, E, \bar{Q}, \bar{U}, \bar{E}$.
The superpotential is
\beq
W = M_Q Q \bar{Q} + M_U U \bar{U} + M_E E \bar{E} + \kappa_U H_u Q\bar{U}
+ \tilde{\kappa}_U H_d \bar{Q} U.
\eeq
The vectorlike masses $M_{Q,U,D}$ can arise by the same
mechanism that generates the $\mu$ term, and are therefore naturally
or order a TeV.

As written, the theory has a global $U(1)$ symmetry that forbids transitions
between these new particles and ordinary quarks and leptons.
If this symmetry is exact, the lightest of the colored
colored particles $Q, U, \bar{Q}, \bar{U}$ is absolutely stable and forms
an $R$-hadron.
(The heavier ones can decay weakly to the lighter ones.)
As discussed above, this $U(1)$ symmetry may be naturally broken by small
effects, so that the long-lived particles are stable only on collider scales.

Assuming large $\tan\be$, the largest correction to the
physical Higgs mass comes from $U$,
and is given by
\beq
\Delta m^2_{h^0} = \frac{N_c}{4\pi^2}
 \ka_U^4 v^2 \sin^4 \beta \log{\frac{M_S^2}{M_F^2}},
\eeq
where $M_F = M_U = M_Q$ is the $U$ fermion mass
and $M_S^2 = M_F^2 + \tilde{m}^2_U$ is the scalar mass.
A positive contribution to the Higgs mass requires $M_F < M_S$,
and therefore the $R$-hadrons are made of the fermions.
The Yukawa coupling of the new fermions to the physical Higgs boson is
approximately 
\beq \label{eq:yU}
y_U = \frac{\ka_U }{\sqrt{2}} U_{Q1} V^\dagger_{\bar{U}1}\sin\beta,
\eeq
assuming that the physical Higgs is aligned with the Higgs VEV.  $U_{Q1}$ is the mixing angle between the lightest mass eigenstate and $Q$ and $V_{\bar{U}1}$ is the mixing between the lightest mass eigenstate and $\bar{U}$.  For $M_Q = M_U$, $U_{Q1} V^\dagger_{\bar{U}1} \sim \frac 12$ for $M_F \gg \kappa_U v \sin\beta$. 

What value should we expect for $\ka_U$?
There is a contribution to the Higgs mass parameter that goes
as $\sim \ka_U^2 M_0^2$, so avoiding fine tuning favors large $\ka_U$.
The Yukawa couplings $\ka_U$, $\tilde\ka_U$
must be $\lsim 1.05$ to avoid Landau poles below the GUT
scale \cite{Martin:2009bg}.
However, larger values of the Yukawa coupling are not necessarily
incompatible with gauge coupling unification, since
it is plausible that the new particles can be composite
at some scale below the GUT scale.
Because these states form a complete GUT multiplet, the composite
dynamics need only be invariant under $SU(5)$ in order to be
compatible with unification.

\section{The Higgsstrahlung Signal}
The 7 TeV LHC is expected to gather 5 fb$^{-1}$ per experiment, and
should be able to discover R-hadrons with  mass up to approximately 800~GeV.
In the event of a discovery, the next task is to
understand the properties of the $R$ hadrons in order to determine
how they fit into a larger theory.
One important clue is the coupling of the Higgs to the $R$-hadrons.
This will be large if the $R$-hadrons get a sizable contribution to their mass
from the Higgs VEV, or equivalently large radiative corrections to the physical
Higgs mass come from the $R$-hadrons.
It is also possible for the $R$-hadron mass to preserve electroweak
symmetry, for example a gluino in supersymmetry.
Observation of Higgs production in association with $R$-hadrons
(``Higgsstrahlung'')
directly measures the coupling of $R$-hadrons to Higgs bosons.
This provides a very clean sample of Higgs bosons, since $R$ hadrons
have no standard-model backgrounds.
The only background to the Higgsstrahlung process is QCD jets produced
in association with the $R$ hadron.
We focus on light Higgs bosons decaying to $\bar{b}b$, and show that
simple jet cuts and a single $b$ tag are sufficient to reduce the
$R$-hadron plus jets background to low levels.
Depending on the $R$-hadron mass, this
can be a discovery mode for a light Higgs boson at the 7 TeV LHC,
and provides a clean sample of Higgs bosons to study
$h \to \bar{b}b$.

\begin{figure}
  \begin{center}
    \scalebox{0.5}{\includegraphics{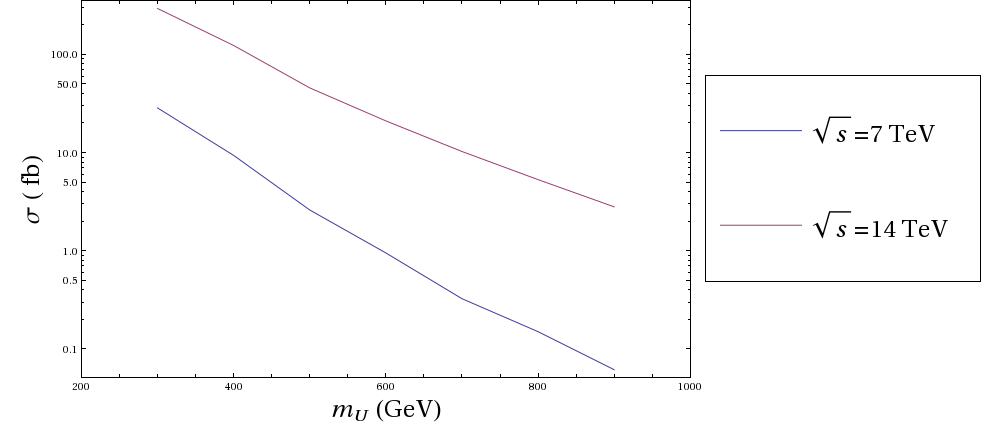}}
  \end{center}
  \caption{The cross section for $\bar{U}Uh$ production for $m_h = 115$ GeV at
  the 7~TeV and 14~TeV LHC.  \label{fig:UUh2LHC}}
\end{figure}

We simulated $\bar{U}Uh$ production in MadGraph \cite{MadGraph}.
Parton-level $\bar{U}Uh$ and $\bar{U}Uhj$ were combined 
to produce a signal sample with up to one additional
hard jet.
We performed a parton-level analysis with a
Gaussian smearing of the jets using the function \cite{JetEnergyRes}
\beq
\frac{\sigma(E_T)}{E_T} = \frac{a}{E_T} \oplus \frac{b}{E_T} \oplus c,
\eeq
for $(a=5.6,b=1.25,c=0.033)$ when $|\eta|<1.4$ and $(a=4.8,b=0.89,c=0.043)$ when $1.4<|\eta|<3.0$.  
We used an assumed $b$-tagging efficiency of $0.3$.
Backgrounds were simulated using AlpGen \cite{ALPGEN}.
We analyzed both the 7 TeV LHC and the 14 TeV LHC.
In both cases, we use $y_U = 1/\sqrt{2}$ in $\Eq{eq:yU}$,
roughly the largest value consistent with perturbative 
coupling constant unification.
Increasing this coupling will improve the situation.
The cross section for $\bar{U}Uh$ production at the LHC is shown in Figure \ref{fig:UUh2LHC} for $m_h = 115$ GeV for the 7 TeV and 14 TeV LHC.

\subsection{7 TeV LHC}

CMS has searched for highly-ionizing tracks from $R$-hadrons with
3.1 pb$^{-1}$ of 7~TeV data \cite{Khachatryan:2011ts},
and for stopped late-decaying $R$-hadrons using
10 pb$^{-1}$ of data \cite{Khachatryan:2010uf}.
These searches are quite model-independent,
and set a lower limit $\gtrsim 275$ GeV.
We will therefore be interested in $R$-hadrons with
masses in the range 300~GeV
to 1~TeV.

For our search it is critical to identify the particles produced in
association with the $R$-hadron,
so we focus on searches with highly ionizing tracks.
An important question is the efficiency with which the $R$-hadrons
will be tagged and reconstructed.
In our benchmark
model the $U$ quark has the same color quantum numbers as the stop,
one of the particles searched for in the CMS search.
We expect that our $R$-hadrons will hadronize in roughly the same way
(given the large theoretical uncertainties in the hadronization),
so we estimate the efficiency by comparing with stops.
The efficiency is most sensitive to the $\beta$ of the produced $R$-hadrons.
We have checked that the $\beta$ distribution for
$R$-hadrons produced in association with a Higgs is similar to the
$\beta$ distribution for pure $R$-hadron production in the relevant
range $\beta \lsim 0.6$ where the search is sensitive.
For stop masses ranging from 300~GeV to 1~TeV
the efficiency to find an event with two $R$-hadrons at CMS is $\simeq 0.2$,
so we assume this value for our signal efficiency.

\begin{figure}
  \begin{center}
    \scalebox{0.5}{\includegraphics{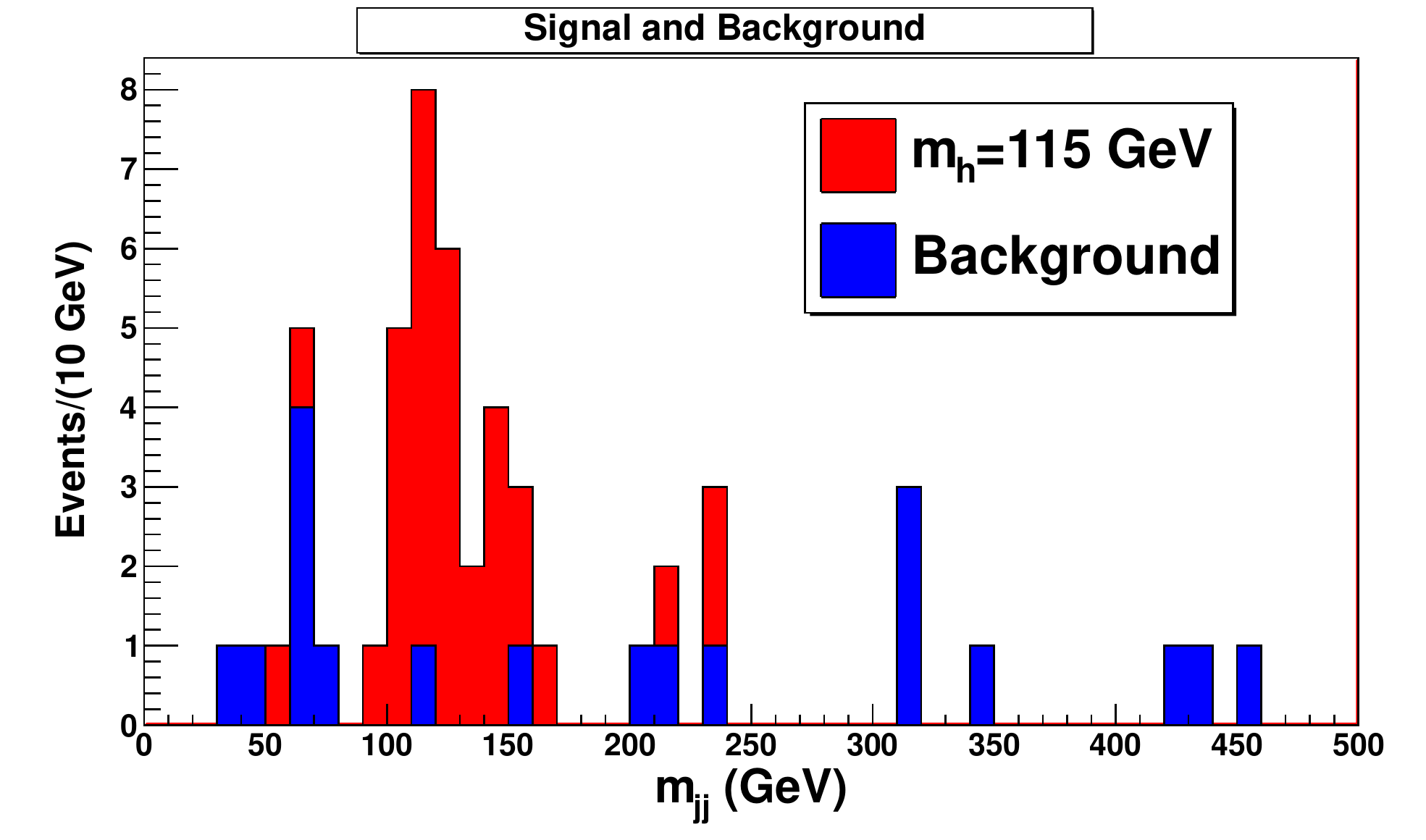}}
  \end{center}
  \caption{10 fb$^{-1}$ of simulated data for the 7 TeV LHC with $m_U = 300$ GeV and $m_h = 115$ GeV, using cuts described in the text.
\label{fig:UUh7}}
\end{figure}

In addition to the $R$ hadron, we require at least 2 jets, each with
$p_T > 50$ GeV and $|\eta| < 2.5$.
We then require that at least one of these jets have a $b$-tag.
We use the events passing these cuts
to form an estimate of the Higgs mass.
In events with 2 $b$-tagged jets we use these to form the invariant
mass, otherwise we combine the $b$-tagged jet with the 
remaining jet with the highest $p_T$.
In a model with $m_U = 300~\mbox{GeV}$, we show the effect of these
cuts on signal and background in Table~\ref{tab:efficiencies7}.
In this model, there are plenty of events to see a Higgs mass peak,
as shown in Fig.~\ref{fig:UUh7}.
For 10~fb$^{-1}$ we estimate
$S/\sqrt{B} \simeq 13$ for the Higgs mass peak defined by
$m_h \pm 25$~GeV.
This will be diluted by a trials factor for Higgs discovery,
but it illustrates that a significant Higgs discovery can be
made early in this channel.

\begin{table}
  \begin{tabular}{|c|c|c|}
    \hline
    & Signal & Background \\
    \hline
    $R$-hadron + 2 jets with $p_T > 50$~GeV, $|\eta|<2.5$
    & 6.2~fb & 220~fb\\
    $\ge 1$ b-tag & 3.2~fb  &2.2~fb\\   
    Higgs mass window $\pm 25$~GeV & 2.5~fb & 0.37~fb \\
    \hline
  \end{tabular}
  \caption{Signal and background rates
  for $m_U = 300$ GeV and $m_h = 115$ GeV at the 7 TeV LHC.
  \label{tab:efficiencies7}}
\end{table}

\subsection{14 TeV LHC}

\begin{figure}
  \begin{center}
    \scalebox{0.5}{\includegraphics{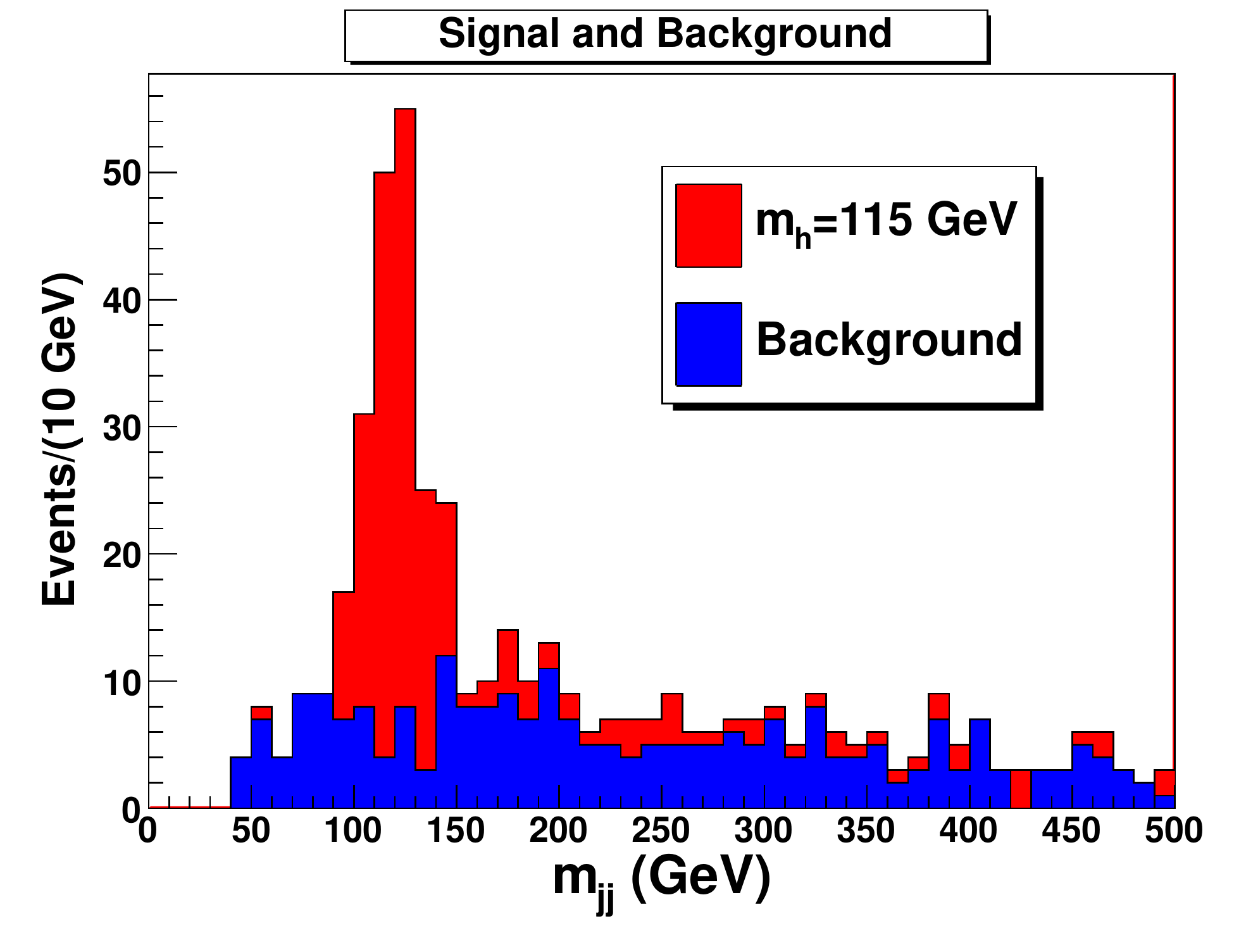}}
  \end{center}
  \caption{100 fb$^{-1}$ of simulated data for the 14 TeV LHC with $m_U = 500$ GeV and $m_h = 115$ GeV, using cuts outlined in Table \ref{tab:efficiencies14}. \label{fig:UUh14}}
\end{figure}

The 14 TeV LHC is expected to be sensitive to $R$-hadron masses up
to about 2~TeV \cite{Fairbairn:2006gg}.
We assume that the signal acceptance for $R$-hadrons is the same
as the 7 TeV LHC, {\it i.e.\/}~$0.2$.
For the increased energy, the background increases at a faster rate
than the signal due to the increased $p_T$ of the jets.
This gives a modest increase to the size of the background relative to the signal, however, $S/B$ can still be made larger than 1.
We find that the discovery reach for this process is rate limited, and we can conservatively hope to probe $m_U < 800$ GeV in the case of $m_h < 125$ GeV with 100 fb$^{-1}$ of data.  Table \ref{tab:efficiencies14} gives and example of the signal cross section after basic cuts, and Figure \ref{fig:UUh14} shows the signal for these same cuts with 100 fb$^{-1}$ of data.

\begin{table}
  \begin{tabular}{|c|c|c|}
    \hline
    & Signal & Background \\
    \hline
    Cross section requiring $R$-hadrons & 8.9 fb & 820 fb\\
    and $h\to\bar{b}b$ or 2 jets &  & \\
    Requiring at least 1 b-tag & 4.5 fb  & 9.5 fb\\   
    Requiring 2 jets with $p_T > 50 $ GeV and $\eta < 2.5$  & 2.2 fb  & 3.2 fb  \\
    In mass window $m_h \pm 25$ GeV & 1.5 fb & 0.42  fb \\
    \hline
  \end{tabular}
  \caption{Example numbers for $m_U = 500$ GeV and $m_h = 115$ GeV at the 14 TeV LHC.  \label{tab:efficiencies14}}
\end{table}

\section{Conclusion}

In this paper we explored the ability of the LHC to discover the Higgs in associated production with R-hadrons.
We found that for a wide range of $R$-hadron masses discovery is possible.
If $R$-hadrons are discovered in the first stage of running at the LHC,
a more careful study should be undertaken to determine the reach of the LHC for
associated Higgs production.
This can determine if a significant part of the $R$-hadron mass arises
from electroweak symmetry breaking, and therefore distinguish between
different models of $R$-hadrons.
Additionally, it gives a  potential clean channel to study $h \to \bar{b}b$.

Note added: While this work was underway, we learned of independent
work by S. Chang, C. Kilic, and T. Okui that considers similar
questions \cite{CKO}.

\section{Acknowledgments}

We would like to thank Aaron Pierce, Tia Miceli, James Dolen, and Max Chertok
for discussions.
ML and DP are supported by DE-FG02-91-ER40674.

\bibliography{XXh}  

\end{document}